\documentclass[twocolumn,notitlepage,aps,prb,10pt,superscriptaddress,floatfix,letterpaper,reprint
]{revtex4-2}	

\usepackage{hyperref}
\usepackage[usenames,dvipsnames]{color}
\usepackage{amsmath}
\usepackage{mathtools}
\usepackage[english]{babel}
\usepackage{amssymb}
\usepackage{graphicx}
\usepackage{epstopdf}
\usepackage[normalem]{ulem}
\usepackage{subfigure}
\usepackage{todonotes}
\usepackage{braket}
\usepackage{bbold}
\usepackage{placeins}
\usepackage[para]{threeparttable}
\usepackage{lipsum}
\usepackage{multirow}
\usepackage{bm}

\definecolor{linkcolor}{HTML}{399B03}
\definecolor{urlcolor}{HTML}{399B03}
\hypersetup{pdfstartview=FitH, linkcolor=linkcolor,urlcolor=urlcolor,colorlinks=true}
\hypersetup{
    unicode=false,          
    pdftoolbar=true,        
    pdfmenubar=true,        
    pdffitwindow=false,     
    pdfstartview={FitH},    
    pdfauthor={},         
    colorlinks=true,        
    linkcolor=blue,         
    citecolor=red,          
}
\begin{document}

\title{Triple excitations in Green's function coupled cluster solver for studies of strongly correlated systems in the framework of self-energy embedding theory}
\author{Avijit Shee}
\affiliation{Department of Chemistry, University of Michigan, Ann Arbor, Michigan 48109, USA}
\affiliation{Department of Chemistry, University of California, Berkeley, California 94720, USA}
\email{ashee@berkeley.edu}
\author{Chia-Nan Yeh}
\affiliation{
 Department of Physics, University of Michigan, Ann Arbor, Michigan 48109, USA
}
\author{Bo Peng}
\affiliation{Pacific Northwest National Laboratory, Richland, Washington, USA}
\author{Karol Kowalski}
\affiliation{Pacific Northwest National Laboratory, Richland, Washington, USA}

\author{Dominika Zgid}
\affiliation{Department of Chemistry, University of Michigan, Ann Arbor, Michigan 48109, USA}
\affiliation{
 Department of Physics, University of Michigan, Ann Arbor, Michigan 48109, USA
}
\email{zgid@umich.edu}

\begin{abstract}


Embedding theories became important approaches used for accurate calculations of both molecules and solids. In these theories, a small chosen subset of orbitals is treated with an accurate method, called an impurity solver, capable of describing higher correlation effects.
Ideally, such a chosen fragment should contain multiple orbitals responsible for the chemical and physical behavior of the compound. Handing a large number of chosen orbitals presents a very significant challenge for the current generation of solvers used in the physics and chemistry community. Here, we develop a Green's function coupled cluster singles doubles and triples (GFCCSDT) solver that can be used for a quantitative description in both molecules and solids. This solver allows us to treat orbital spaces that are inaccessible to other accurate solvers. At the same time, GFCCSDT maintains high accuracy of the resulting self-energy. Moreover, in conjunction with the GFCCSD solver, it allows us to test the systematic convergence of computational studies.
Developing the CC family of solvers paves the road to fully systematic Green's function embedding calculations in solids.
In this paper, we focus on the investigation of GFCCSDT self-energies for a strongly correlated problem of SrMnO$_3$ solid. Subsequently, we apply this solver to solid MnO showing that an approximate variant of GFCCSDT is capable of yielding a high accuracy orbital resolved spectral function.

\end{abstract}
\maketitle


Realistic molecules and solids often display features allowing them to be described by embedding methods. In these realistic problems, the orbital space usually can be divided into two parts: a fragment (system) requiring an accurate description and an environment that can be represented only approximately. Usually, the embedding construction is done in such a way that first the entire problem is evaluated via a low-scaling method yielding a low-energy impurity Hamiltonian for the system/chosen fragment. Afterwards, the low-energy system Hamiltonian is treated with a more accurate non-perturbative method.

Such an embedding construction is not only applicable for solids and molecules containing transition metals with $d$- and $f$-shells, where the local impurity is small and composed of $d$- and $f$-orbital, but also for dispersion dominated solids, for example, molecular crystals, two-dimensional layered materials. In these materials, the local impurity may be large and may involve a large number of orbitals, often a big supercell \cite{Gruneis_JCP2021}.

Both self-energy embedding theory (SEET) \cite{Zgid_2011,Tran_jcp_2015,Tran_jctc_2016,Lan17,Tran_Shee_2017, Tran_useet} and dynamical mean field theory (DMFT) \cite{Georges96, Kotliar06, LDAplusDMFT_Anisimov1997,Held2006} are Green's function-based quantum embedding theories, which provide a general framework for the treatment of embedding problems. 
SEET was employed in molecular problems~\cite{Zgid_2011,Tran_jcp_2015,Tran_jctc_2016,Lan17,Tran_Shee_2017, Tran_useet} as well as strongly correlated solids~\cite{Iskakov20, YehPRB2021}.

The choice of the non-perturbative method employed in the treatment of the chosen fragment (or the impurity problem) is often challenging for both SEET and DMFT. This is because this method (frequently called a solver) should treat the impurity problem accurately. In particular, for strongly correlated problems, it is hardly possible to treat impurities that are larger than 16 orbitals. 

One of the most widely employed solvers, continuous-time quantum Monte Carlo (CTQMC) \cite{Gull2011}, is applicable only for a few sites, limited to Hubbard-type interactions, and fraught with fermionic sign problem. Another method, Exact Digaonalization (ED) \cite{Cafarrel_ED_PRL1994}, which requires discretized impurity bath can treat generalized interaction, but also limited to very few sites because of its exponential scaling. In order to alleviate those limitations many truncated wave-function  based solvers, for example, DMRG-based \cite{Wolf_PhysRevX2015}, selected and truncated CI-based \cite{Carlos_PhysRevB_2019, Zgid2012} solvers have been proposed.  Usually these truncated solvers are employed for those problems, where a larger number of sites is considered and the accuracy can be somewhat sacrificed in comparison to ED.           


 We have chosen the coupled cluster \cite{Cizek_CC,Paldus_Cizek} (CC) hierarchy of methods as solvers for solving the impurity problems. CC methods are very successful in quantum chemistry \cite{Bartlett07}.
 Recently, we have implemented a Green's function mapping of the CC wave function for use in embedding frameworks \cite{Shee2019}. In our previous works, we have employed coupled cluster singles doubles truncation (CCSD) as an impurity solver called here GFCCSD. Similar solver has been developed by Zhu \emph{et. al.} \cite{Zhu_PRB19, Zhu_DMFT_PRX21} in the dynamical mean field theory (DMFT) context. However, for a truncated solver it is not guaranteed to be successful at all interaction strengths. Therefore, in our previous works, both with periodic solids \cite{Yeh_SheePRB21} and molecules \cite{SheeJCTC2022} we have explored strengths and weaknesses of the GFCCSD solver. We observed that GFCCSD (i) frequently failed to identify a correct ground state of the impurity problem (ii) was reasonably successful when  the number of impurity orbitals was small (iii) with a limited truncation at the SD level is hard to be completely decisive if the convergence with respect to excitation level was reached.

    


Therefore, in this work we will improve upon the GFCCSD solver by including triple excitations. For CC hierarchy, the inclusion of higher rank excitations provides a much faster convergence in comparison to other wave function methods, such as, configuration interaction (CI) and many body perturbation theory (MBPT) \cite{KallayJCP2001}. Moreover, historically the inclusion of triple excitations in schemes such as CCSD(T) provided unprecedented accuracy in molecular problems.
With the inclusion of full triple excitations, the computational complexity increases from $\mathcal{O}(N^6)$ to $\mathcal{O}(N^8)$. Therefore, it is important to make approximations so that the steep scaling of a triples calculation can be reduced. We have proposed such an approximate scheme in the next section that has the computational complexity $\mathcal{O}(N^7)$. To our knowledge, this is the first reported GFCCSDT implementation including full triples, which has been used in a quantum embedding context. A GFCC scheme including approximate triples excitation has been reported previously by Peng and Kowalski \cite{Peng_JCP2018}.    

The GFCC solver with triple excitations can be essential for quantitatively accurate embedding calculations. This is because in realistic applications of embedding methods, we often want to achieve a systematic improvement/convergence of the observables (total energy, spectral function etc.) in terms of various computational choices. Two such choices relevant only for the embedding part of a calculation are: (a) the accuracy of the bath representation in the impurity problem ; (b) the number of orbitals in the local problem chosen for embedding. 

The GFCCSD solver with triple excitations will allow us to employ a very large number of bath orbitals, thus ensuring accurate bath representation and a convergence with respect to the number of bath orbitals. 
Moreover, it is often difficult to choose a local fragment which is optimal for the description of the chosen system. In GFCCSDT,  we can systematically increase the number of orbitals that are chosen while still maintaining high accuracy of calculations.
Both of these advantages of the newly developed solver will allow for quantum embedding calculations to become systematically improvable and highly quantitatively accurate. 

CC is a many-body theory based on the exponential parametrization of the ket wave function $|\Psi \rangle = e^{\hat{T}} |\Phi \rangle $, where $|\Phi\rangle$ is a reference mean-field wavefunction. In this paper, we are using a unrestricted Hartree Fock (UHF) determinant $|\Phi_{\text{UHF}}\rangle$ as the reference wave function. $T$ is a cluster operator inducing various h-p excitation from the reference wave function $|\Phi\rangle$. 

\begin{equation}
\hat{T} = \sum_{ai} t^a_i \{a_a^{\dag} a_i \} + \sum_{\mathclap{a>b, i>j}} t^{ab}_{ij} \{ a^{\dag}_a a^{\dag}_b  a_j^{\phantom{\dag}} a_i^{\phantom{\dag}}\} 
\end{equation}

In CC, the bra wave function is not an adjoint of the ket because $e^{\hat{T}}$ is not a unitary operator. In general, the choice of the bra state is non-unique in CC. Here, we define $\langle \Psi_L| = \langle \Phi| (1+\hat{\Lambda}) e^{-\hat{T}}$ through the de-excitation operator $\Lambda$ as the bra state \cite{ARPONEN_1983}. This definition ensures that the bra and ket states are biorthonormal. A CC wave function is mapped to a corresponding Green's function \cite{Shee2019,NooijenIJQC1992,KowalskiJCP2014,ZhuPRB19} by the following definition:

\begin{align}
G^{CC}_{pq} (\omega) = {} & \langle \Phi | (1+ \hat{\Lambda}) \overline{a_p^\dagger} \frac{1}{\omega + \overline{H} - i\eta } \overline{a}_q| \Phi \rangle \nonumber \\ 
     {} & +  \langle \Phi | (1+ \hat{\Lambda}) \overline{a}_p \frac{1}{\omega - \overline{H} + i\eta}  \overline{a_q^\dagger} | \Phi \rangle \label{Eq:ccgf2},
\end{align}
where $\overline{a}_p = e^{-\hat{T}} a_p e^{\hat{T}}$ and $\overline{a_p^\dagger} = e^{-\hat{T}} a_p^\dagger e^{\hat{T}}$. The creation (annihilation) operator is  $a^{\dag}_{p}$ ($a_{p}$) operating on a single-particle state $p$. The transformed Hamiltonian is defined as $\overline{H} = e^{-\hat{T}} H e^{\hat{T}} - E_{gr}$. The UCC ground state energy is denoted as $E_{gr}$. 
While CC is equivalent to ED if no truncation is made in the rank of the cluster operator $\hat{T}$, in practical implementations, the rank of cluster operator is always truncated to minimize computational cost.  In this work, we employ the singles-doubles-triples (SDT) approximation, resulting in the following operators $\hat{T} = \hat{T}_1 + \hat{T}_2 + \hat{T}_3$ ; $\hat{\Lambda}$ = $\hat{\Lambda}_1$ + $\hat{\Lambda}_2$ + $\hat{\Lambda}_3$ for the ground state problem. In a subsequent step, we tridiagonalize $\overline{H}$ in the space of $(N+1)$ and $(N-1)$ electronic wave functions, using the Lanczos algorithm. The $(N+1)$ and $(N-1)$ electronic states are spanned by all the functions generated by $\hat{R}$ and $\hat{L}$ operators defined as 
\begin{equation}
\hat{R}^{{\rm (N+1)}} = \sum_{a} r^a \{a_a^{\dag} \} + \sum_{\mathclap{a>b, i}} r^{ab}_{i} \{ a^{\dag}_a a_b^{\dag} a_i^{\phantom{\dag}} \} + \sum_{\mathclap{\substack{a>b>c,\\i>j}}} r^{abc}_{ij} \{ a^{\dag}_a a^{\dag}_b a^{\dag}_c   a_j^{\phantom{\dag}} a_i^{\phantom{\dag}}\}
\end{equation}

\begin{equation}
\hat{L}^{{\rm (N+1)}} = \sum_{a} l_a \{a_a^{\phantom{\dag}} \} + \sum_{\mathclap{i,a>b}} l_{ab}^{i} \{ a^{\dag}_i a_b^{\phantom{\dag}} a_a^{\phantom{\dag}} \} + \sum_{\mathclap{\substack{i>j,\\a>b>c}}} l_{abc}^{ij} \{ a^{\dag}_i a^{\dag}_j   a_c^{\phantom{\dag}} a_b^{\phantom{\dag}} a_a^{\phantom{\dag}} \}
\end{equation}
 
\begin{equation}
\hat{R}^{{\rm (N-1)}} =  \sum_i r_i \{a_i^{\phantom{\dag}}\} + \sum_{\mathclap{i>j,a}} r_{ij}^{a} \{a^{\dag}_a a_j^{\phantom{\dag}} a_i^{\phantom{\dag}}\} + \sum_{\mathclap{\substack{i>j>k,\\ a>b}}} r_{ijk}^{ab} \{a^{\dag}_a a^{\dag}_b a_k^{\phantom{\dag}} a_j^{\phantom{\dag}} a_i^{\phantom{\dag}}\} 
\end{equation}
\begin{equation}
\hat{L}^{{\rm (N-1)}} =  \sum_i l^i \{a^{\dag}_i\} + \sum_{\mathclap{i>j,a}} l^{ij}_{a} \{a^{\dag}_j a^{\dag}_i a_a\} + \sum_{\mathclap{\substack{i>j>k,\\ a>b}}} l^{ijk}_{ab} \{a^{\dag}_i a^{\dag}_j a^{\dag}_k a_b a_a\}
\end{equation}
and acting on  $e^{\hat{T}} |\Phi \rangle $ and $\langle \Phi | e^{-\hat{T}}$, respectively.
Both $\hat{L}$ and $\hat{R}$ operators are truncated at the level of triples. 

The computational scaling of the evaluation of the ground state problem with SDT approximation scales as $n_b^8$, whereas the tridiagonalization step scales as $n_b^7$, where $n_b$ is the number of basis functions. The maximum memory requirement for the SDT truncation is due to storing $T_3$ and $\Lambda_3$ amplitudes. In order to make both computational scaling and memory requirements manageable, methods that include triples are often employed with various approximations that effectively lower the compuational and memory scaling. In this work, we have used an approximation proposed by Hirata \textit{et. al.} \cite{Hirata_cpl2000} in their EOM-CCSD ($m$,$n$) work, where $m$ and $n$ represents the rank of excitations in the ground state and in the excited state problem, respectively. Krylov and Slipchenko \cite{slipchenko2005} later analyzed the strength and weaknesses of such an approximation in the EOMCC context and discussed the lack of size intensivity of the excited state variant. 
It has been further shown by Peng and Kowalski\cite{Peng_JCP2018} that in order to maintain size extensivity of the resulting Green's function, the rank of n should not exceed by more than one in comparison to m. 

In this spirit, we choose the ground state cluster operator $\hat{T}$ and $\hat{\Lambda}$ operator up to rank 2. For the $(N+1)$ and $(N-1)$ electronic problems, we choose the $\hat{R}$ and $\hat{L}$ up to excitation rank 3. This approximation not only allows us to avoid the steep $n_b^8$ computational scaling, but also it allows for memory savings - we no longer need arrays of dimension ($n_o^3$$n_v^3$) ($n_{o}$ and $n_{v}$ stand for number of occupied and virtual orbitals, respectively) for $T_3$ and $\Lambda_3$. The maximum size of necessary arrays for $R_3$ and $L_3$ is now $n_o^3 n_v^2$/$n_o^2 n_v^3$ for $(N-1)$/$(N+1)$ problems, respectively. 

We can rationalize the EOM-CCSD ($m$,$n$) approximation by using a biorthogonal projection operator ($\sum_k \hat{R}_k e^{\hat{T}} |\Phi_0\rangle\langle \Phi_0 | \hat{L}_k e^{-\hat{T}}$) for the $(N-1)$ states in Eq. \ref{Eq:ccgf2}, resulting in the following expression:
\begin{equation}
    G_{pq}^{CC,N-1} = \sum_k \frac {Y_{pk}^\dag X_{kq}}{\omega + \Delta E_k - i\eta}
\end{equation}
where,
\begin{equation}
    Y_{pk}^\dag = \langle \Psi_L | a_p^\dag | \Psi_k^{N-1} \rangle ; \quad X_{kq} = \langle \Psi_k^{N-1} | a_q | \Psi \rangle
\end{equation}
and $\Delta E_k$ corresponds to the electron detachment energy, when $R_k$ and $L_k$ are the ket and bra eigenvectors of the $N-1$ electronic problem.
With GFCCSDT(2,3) approximation,  $| \Psi_k^{N-1} \rangle$ spans much larger space since it contains 3h-2p states as well. Likewise, $| \Psi_k^{N+1} \rangle$ also includes 3p-2h states, thus with GFCCSDT(2,3) approximation, we span larger space than GFCCSD alone. 
This space is necessary to describe both the ionization and attachment processes that affect the CCSD wavefunction.

 The inversion present in Eq. \ref{Eq:ccgf2}  is carried out using a continued fraction formula \cite{Shee2019}. Consequently, the computational scaling of our approach is independent of the size of the frequency grid since the Green's function at any arbitrary frequency (both on the real and imaginary frequency axis) can be evaluated using the continued fraction formula.

In order to assess the accuracy of the solvers based on GFCCSDT and GFCCSDT(2,3) developed above, we considered a cubic perovskite, SrMnO$_{3}$, which has a $G$-type antiferromagnetic (AFM) ordering at low-temperature and paramagnetic (PM) ordering at room temperature. 
Ne\'{e}l temperature for SrMnO$_3$ has been reported to be $\sim$ 233$-$260 K~\cite{Negas70,Takeda74}. 
In this work, we focus on the high-temperature PM insulating phase, for which several photoemission experimental data ~\cite{Abbate92,Saitoh95,SrMnFeO3_expt_Kim10} are available, and they predict a band gap of 1.0 to 2.3 eV. 


In our previous work~\cite{YehPRB2021}, we have shown that sc$GW$ predicts an incorrect metallic PM phase for this system. Several other more sophisticated studies, such as LDA+DMFT~\cite{Dang14,Chen14,Bauernfeind18} and multitier $GW$+EDMFT~\cite{GW_EDMFT_PRM_Philipp20}, fail to open the gap in the PM phase as well. Recently, our group has reported a study of the same PM phase, where SEET($GW$/ED) was successfully employed to open a gap with an outer-loop self-consistency~\cite{YehPRB2022}. 
In that study, from the evaluated orbital resolved spectra, we observed that  Mn:$3d$ and O:$2p$ orbitals have dominant contributions to bands near the Fermi level. In the SEET($GW$/ED) study, we created a series of impurity problems built from Mn:$3d$ and O:$2p$ orbitals. The self-energies obtained using ED for these impurities were used to correct the $GW$ self-energy during the self-consistency SEET cycle. 

In a later study \cite{Yeh_SheePRB21}, we have used the same SrMnO$_{3}$ PM phase to assess the accuracy of the GFCCSD solver for the impurity problems. It revealed that for some of the impurity choices, GFCCSD predicts similarly qualitative results as ED. For example, SEET($GW$/GFCCSD), employing GFCCSD as a solver for the Mn:$t_{2g}$ impurity, opens a gap yielding quantitatively correct spectral functions. 

However, for some other impurity choices GFCCSD was not as successful, resulting in quantitatively incorrect self-energies that affected the final spectral functions. For example, with Mn:$e_g$ and O:$p_\pi$ impurities, SEET with  GFCCSD as a solver predicts a spurious peak near the Fermi region \cite{Yeh_SheePRB21} yielding incorrect spectral functions.

In this paper, we will investigate whether inclusion of triples improves upon the GFCCSD self-energies. This improvement will be analyzed by examining self-energies of various impurities of Mn:$3d$ and O:$2p$ orbitals. Subsequently, we will investigate the effect of the improved impurity self-energies built from Mn:$3d$ and O:$2p$ orbitals as shown in Table \ref{tab:imp_choice} on the full SEET($GW$/GFCCSDT) calculations.

We have assembled all the integrals for those impurity problems in an open access repository \cite{shee_avijit_2022_7329098}.

We have reported two sets of results. In the first set to enable comparison between the ED, GFCCSDT, and GFCCSDT(2,3) self-energies, we were able to carry out these calculations for A, B, D and E impurities. For the remaining impurities, F,G, and H, the ED results are impossible to obtain because of the large number of orbitals (system+bath) in such impurity problems. For these impurities the total number of orbitals exceeds 16. Note also that for all the impurities reported here, the real part of $\Sigma$ is completely dynamic (frequency dependent), since we have subtracted all the static self-energy contributions. 

\begin{figure*}
    \centering
    \begin{tabular}{cc}
    \includegraphics[width=0.45\textwidth]{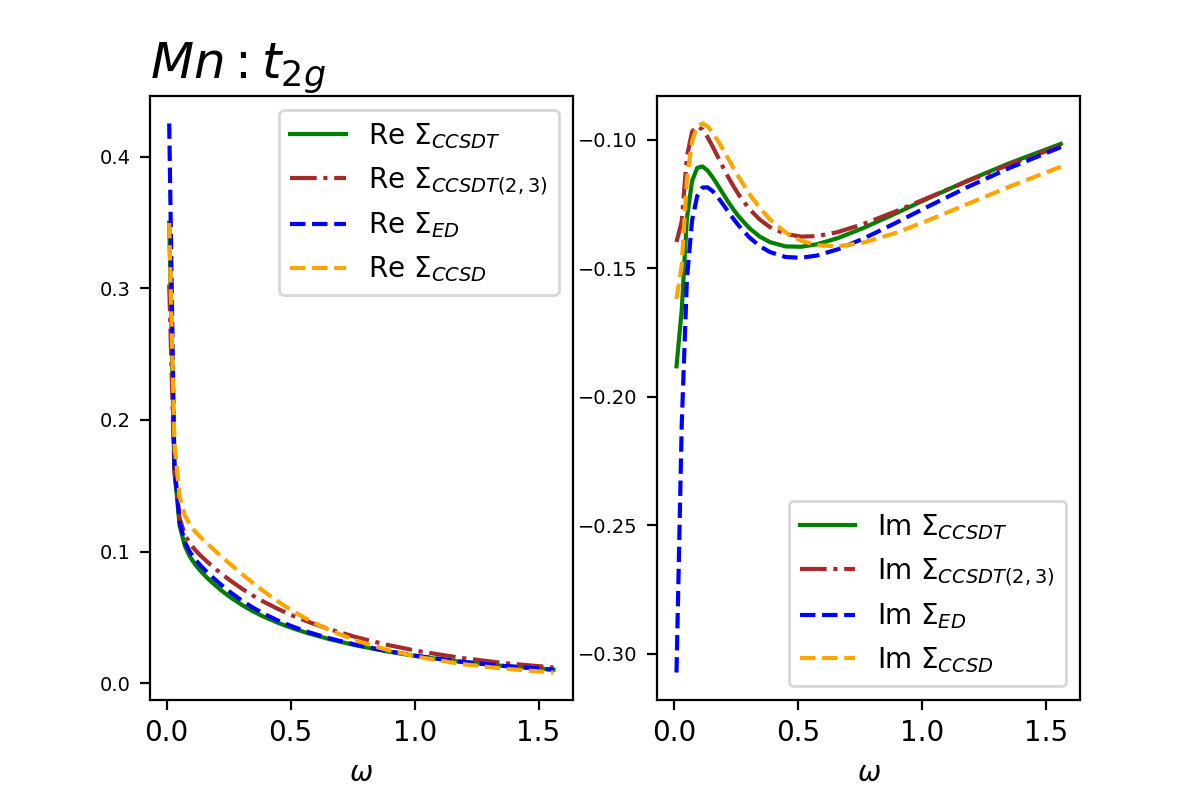} &
    \includegraphics[width=0.45\textwidth]{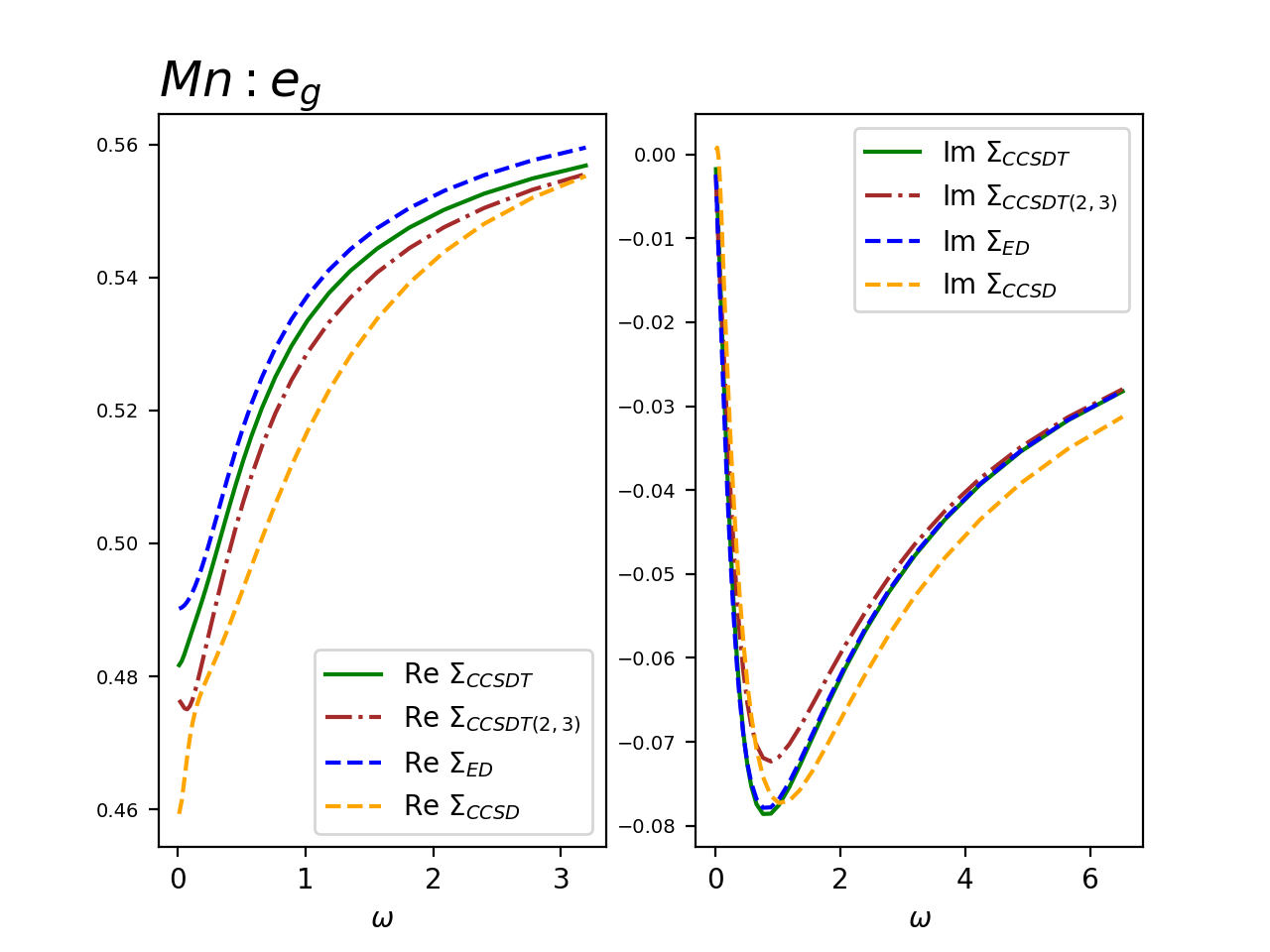}  \\
    \includegraphics[width=0.45\textwidth,height=0.305\textwidth]{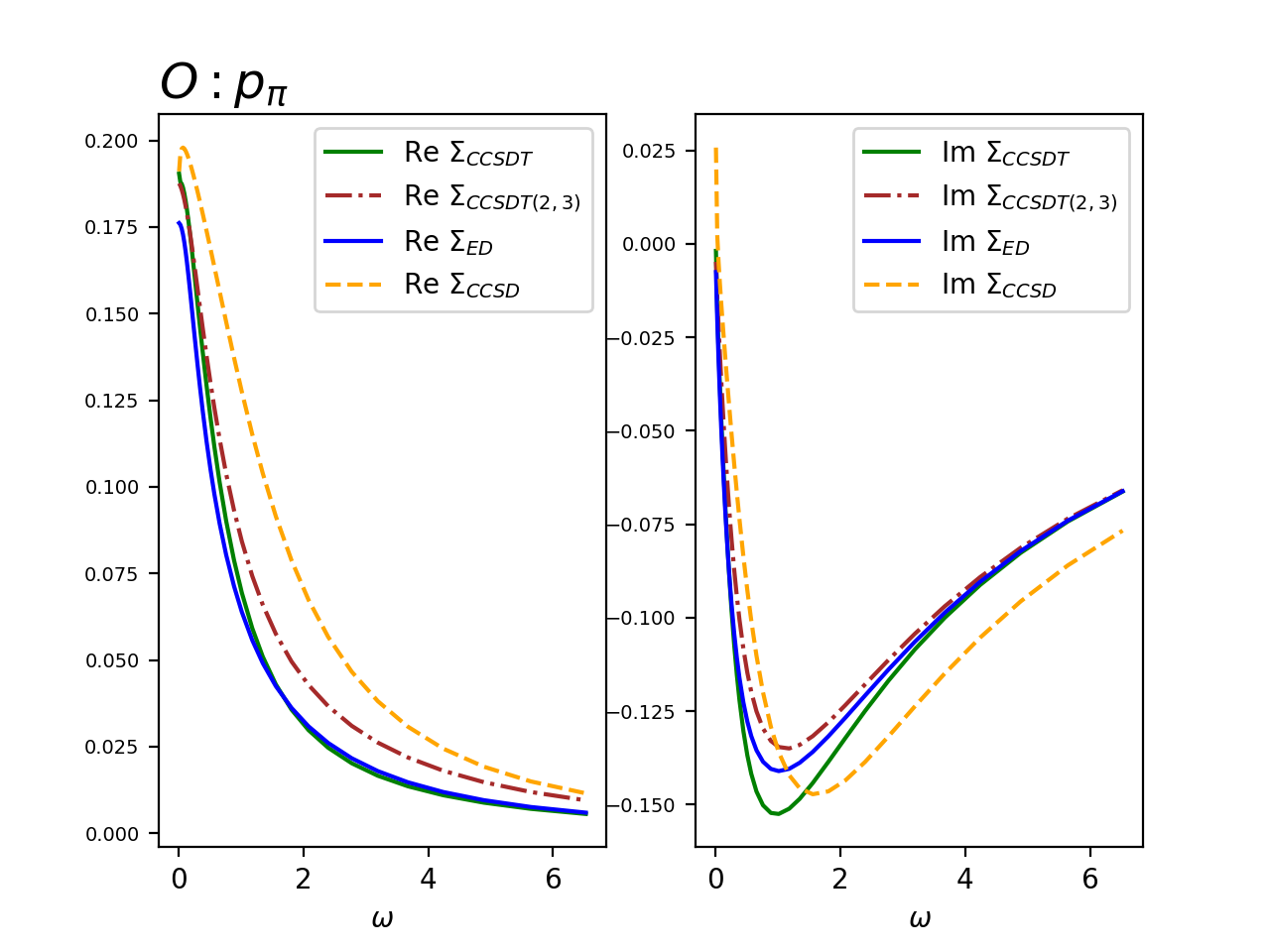} &
    \includegraphics[width=0.45\textwidth]{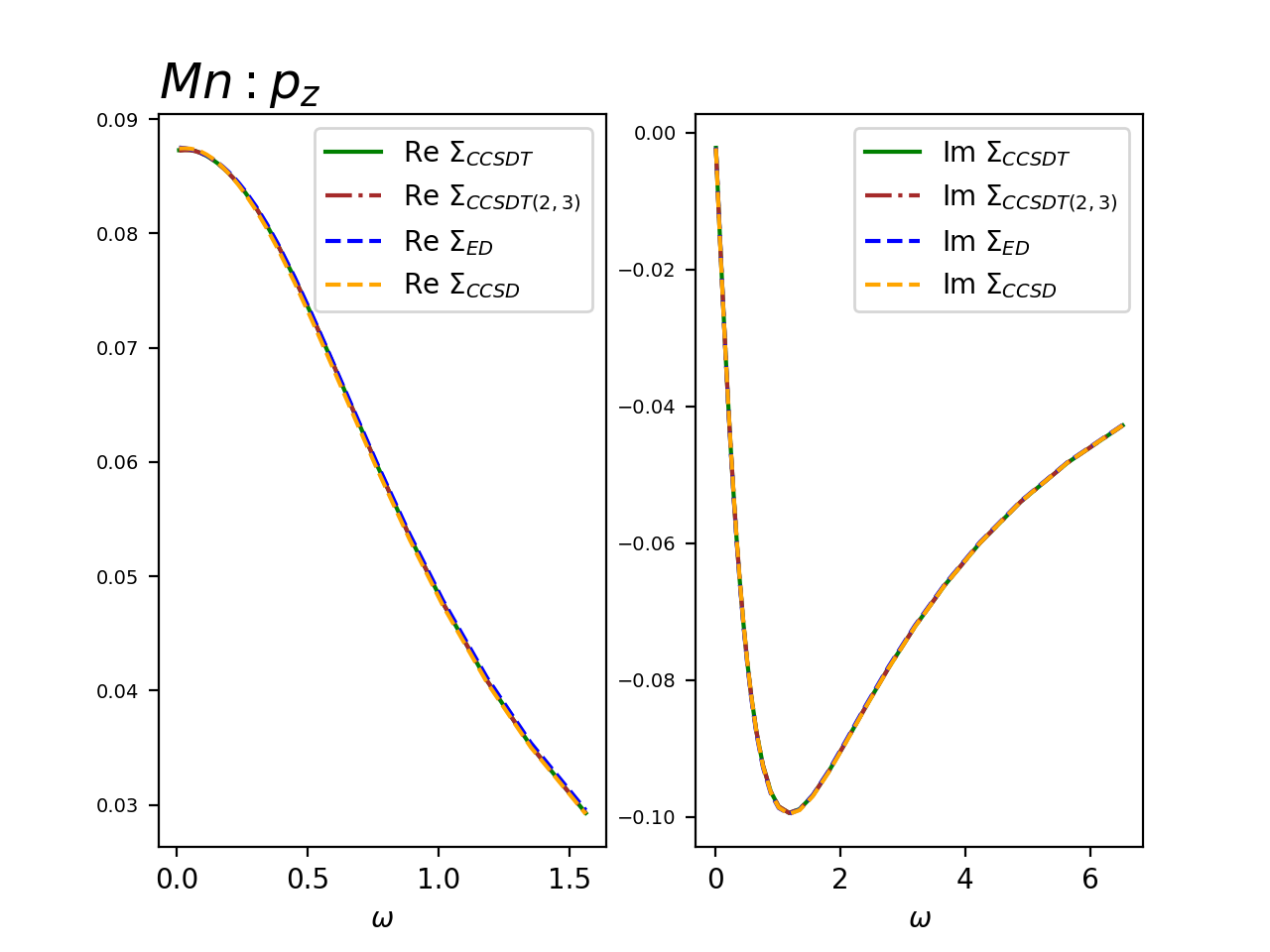} 
         
    \end{tabular}
    \caption{Real and imaginary components of self-energy is compared with various impurity solvers based on GFCC. Impurities are constructed with SrMnO$_3$ solid. The impurities considered are A: Mn: $t_{2g}$, B: Mn:$e_g$ (top panel) and D: O:p$_\pi$, E:  O:$p_z$ (bottom panel). Note that $p_z$ and $p_\sigma$ refers to the same orbital as used in the text. Reference data is obtained from ED calculation.}
    \label{fig:ed_ref}
\end{figure*}

The first set of results in which we are reporting self-energies for impurities A, B, D, E are shown in Fig. \ref{fig:ed_ref}. For all the impurities other than O:$p_\sigma$, we observe quantitative differences when their self-energies at the GFCCSD level are compared to the ED ones. Moreover, the imaginary parts of the GFCCSD self-energies for Mn:$e_{g}$ and O:$p_\pi$ display different minima at the frequency axis in comparison to ED. After adding triples correction with our low-scaling approximated GFCCSDT(2,3) scheme, we can see that they reproduce ED self-energies with better accuracy, especially there is a noticeable improvement in the low frequency regime of Im$\Sigma$ and in Re$\Sigma$. Furthermore, when employing the full GFCCSDT scheme, the self-energy  for Mn:$t_{2g}$ and Mn:$e_g$ impurities is further improved, as we expect. However, for O:$p_\pi$ impurity, Im $\Sigma$ is much better reproduced around minimum with GFCCSDT(2,3) in comparison to GFCCSDT. 
Another crucial observation is that in both for Mn:$e_g$ and O:$p_\pi$ impurities, GFCCSD produces a non-causal self-energy since the GFCCSD Im $\Sigma$ is positive. (A causal, physical diagonal parts of Im $\Sigma$ are always negative). When triples corrections are employed, in both GFCCSDT and GFCCSDT(2,3), such a behaviour is alleviated.  

\begin{figure}
    \centering
    \includegraphics[width=0.47\textwidth]{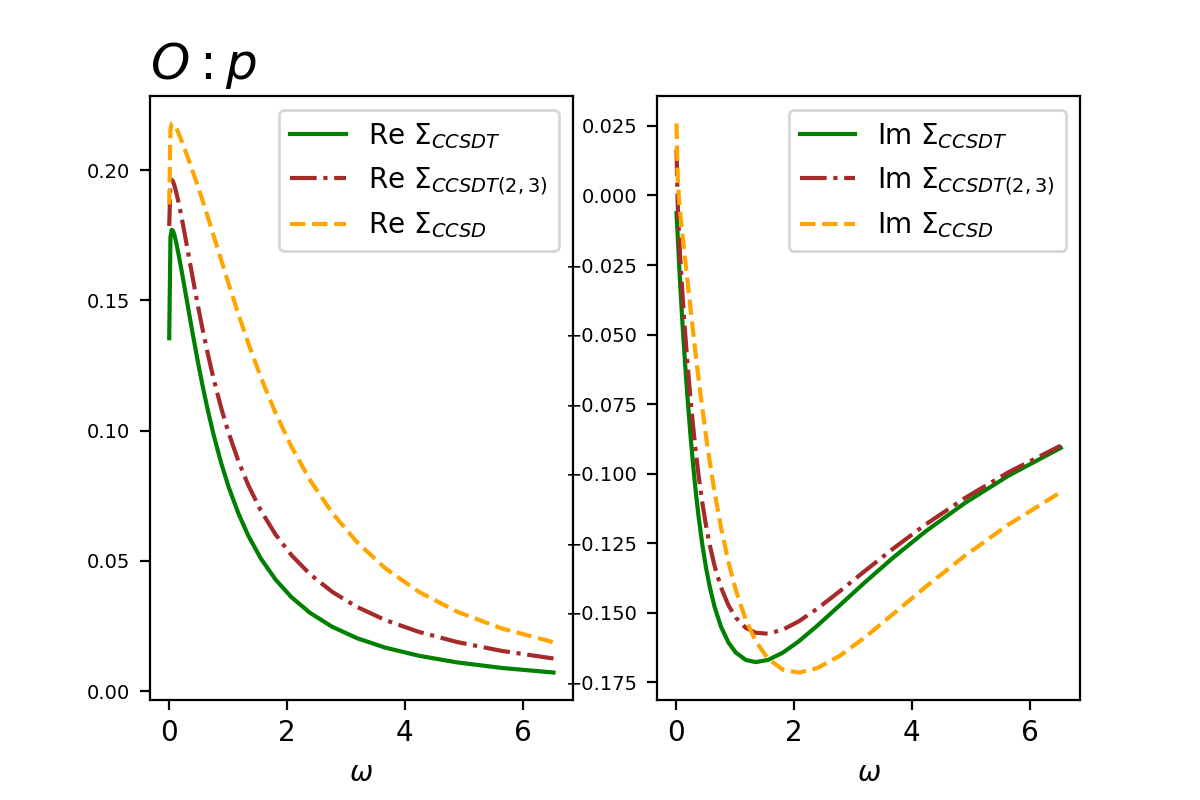}
    \includegraphics[width=0.47\textwidth]{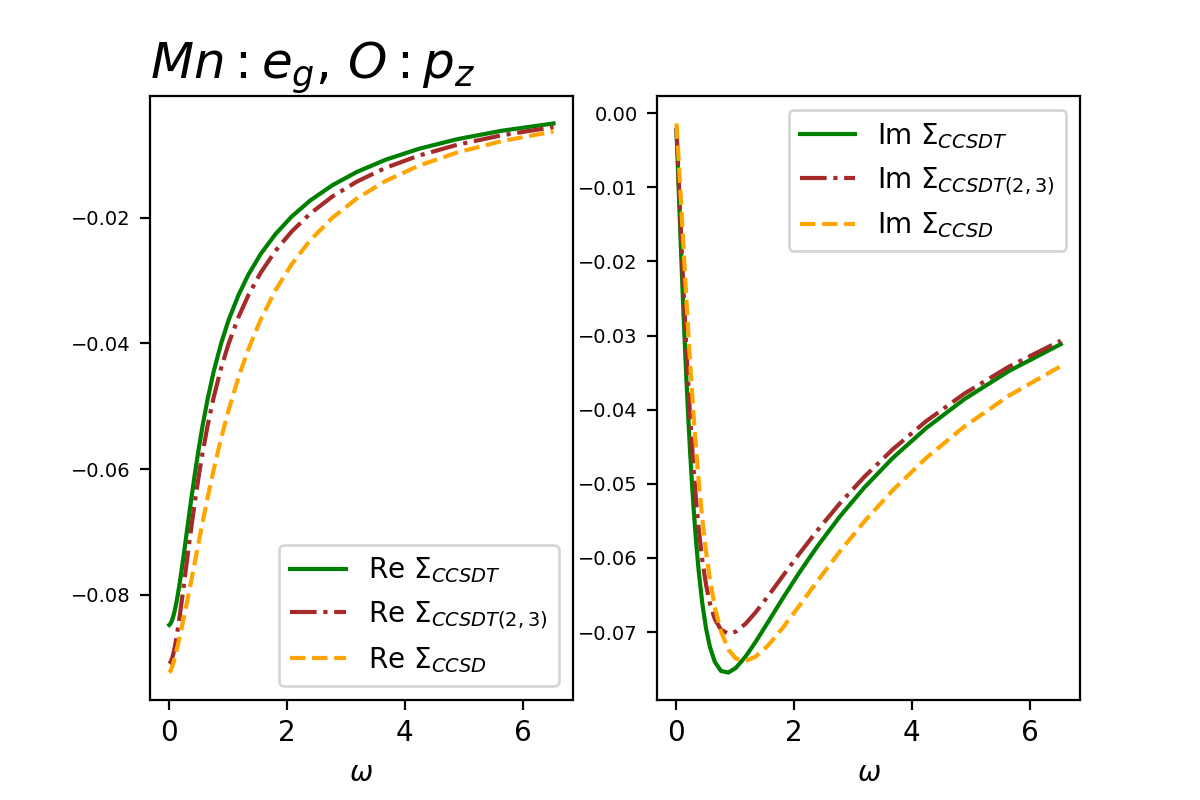}
    \includegraphics[width=0.47\textwidth]{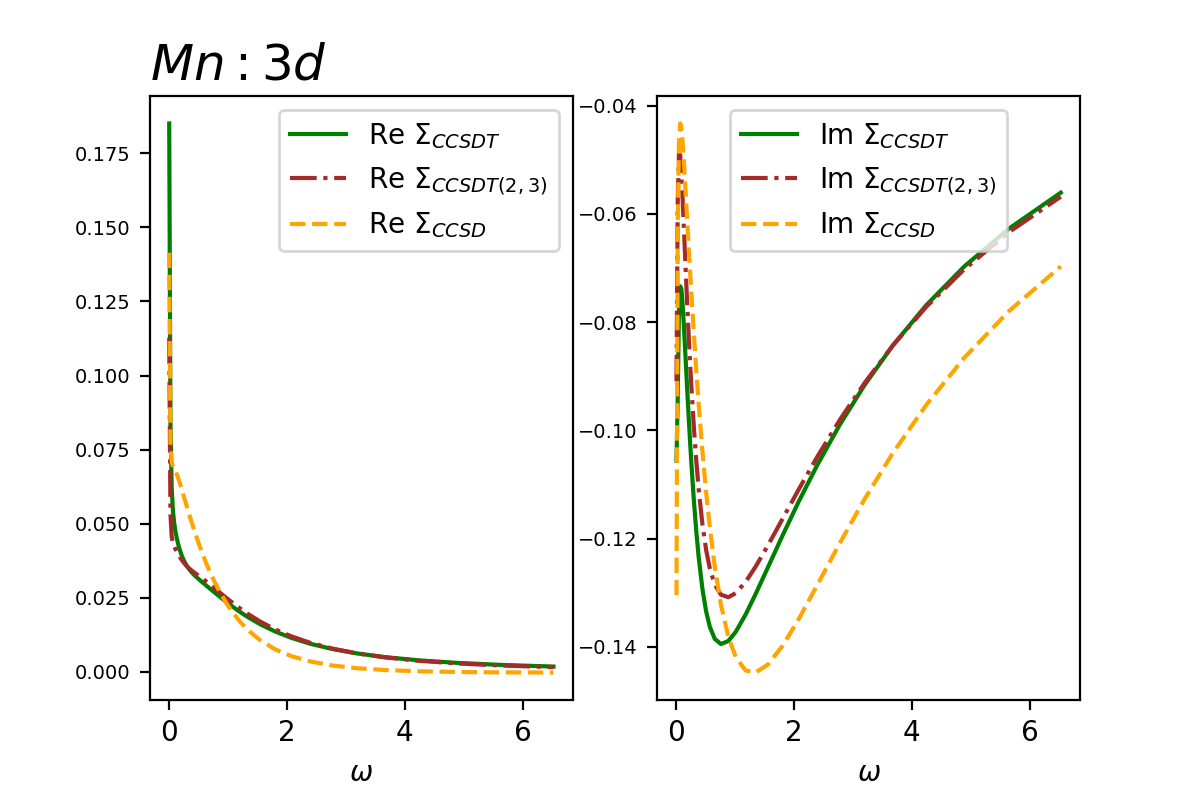}
    \caption{Comparison of real and imaginary self-energy components obtained with various variants of GFCC methods for impurity problems constructed in SrMnO$_3$ solid. O:$p_\pi$ of O:$p$ or impurity F (top panel); Mn:$e_g$ of Mn:$e_g$ O:$p_z$ or impurity G (middle panel), Mn:$t_{2g}$ of Mn:$3d$ or impurity C (bottom panel).  }
    \label{fig:ccsdt_ref}
\end{figure}

The next set of impurities, C, F and G, is created by combining smaller impurities together.  For each of these impurities, this results in a larger size of the combined orbital space (system+bath). Consequently, we are no longer able to compare the resulting self-energies against ED. Therefore, we consider the GFCCSDT self-energy as our reference. Here, we aim to investigate the qualitative and quantitative behaviour of the GFCCSDT(2,3) self-energy. 
For O:$2p$ impurity, we observe from Fig. \ref{fig:ccsdt_ref} that Re $\Sigma$ behaves qualitatively similar for all the methods considered here and GFCCSDT is very similar to GFCCSDT(2,3). However, for Im $\Sigma$ we observe non-causal contributions, both for GFCCSD and GFCCSDT(2,3) methods. This is reminiscent of the O:$p_\pi$ smaller impurity, but unlike for that impurity, here GFCCSDT(2,3) failed to yield a causal self-energy. GFCCSDT, on the other hand does not suffer from this problem. Therefore, by comparing a smaller O:$p_\pi$ and a combined O:$p$ impurity, we conclude that a better performance of GFCCSDT(2,3) for O:$p_\pi$ was probably fortuitous (perhaps due to some error cancellation). For the other larger impurities considered, that is for Mn:$3d$ and Mn:$e_g$ + O:$p_z$, we observe that GFCCSDT and GFCCSDT(2,3) follow each other closely in most parts except showing some quantitative difference at low frequency regime of Im $\Sigma$. GFCCSD, on the other hand differs from other schemes containing triples, both in terms of the position of the minima and absolute values.


\begin{table}

\caption{Different choices of impurity problems for SrMnO$_3$}

\begin{tabular}{|c|c|}
\hline 
Imp & Composition\\
\hline 
\hline 
A & Mn:$t_{2g}$\\
\hline 
B & Mn:$e_g$\\
\hline 
C & Mn:$t_{2g}, e_g$\\
\hline 
D & O:$p_\pi$ ($p_x$, $p_y$) \\
\hline 
E & O:$p_\sigma$ ($p_z$)\\
\hline 
F & O:$p_\sigma, p_\pi$\\
\hline 
G & Mn:$e_g$, O:$p_\sigma$\\
\hline 
H & Mn:$t_{2g}$, O:$p_\pi$\\
\hline 
\end{tabular} \label{tab:imp_choice}

\end{table}

We analyzed the accuracy of total energy obtained from various GFCC variants while taking into account a lack of self-consistent nature of the CC theory. 

Fully self-consistent Green's function methods such as Green's function self-consistent second order (GF2) or GW are approximations of the exact Luttinger–Ward functional $\Phi$.
Diagrammatically, such a functional is the sum of all bold, closed, two-particle irreducible Feynman diagrams (also called “skeleton” diagrams). In such methods $\frac{\partial \Phi}{\partial G}=\Sigma$ and Green's function, $G(\Sigma, v)$, is a functional of the self-energy and bare Coulomb interactions. Consequently, the self-energy and Green's function are obtained self-consistently.

The GFCC Green's function is not constructed as a self-consistent functional of $G$ and $\Sigma$.
We will analyze the error introduced due to the non-self-consistent nature of the self-energy obtained from all the GFCC variants described here. In order to do so, we will consider one traced quantity, namely the total energy for the impurity. We have evaluated the total energy in two different ways.

First, we evaluated it directly from the CC wave function as
\begin{equation}
    E_{1b}^{CC} = \frac{1}{2}\gamma_p^q (h^p_q + F^p_q) ; \quad E_{2b}^{CC} = \frac{1}{4} \Gamma_{pq}^{rs} V^{pq}_{rs}, \label{eq:wf-energy}
\end{equation}
where, $h_p^q$, $F_p^q$, $V^{pq}_{rs}$ are the bare one-electron, Fock and bare two-electronic Hamiltonian matrix elements in the molecular orbital (MO) basis, and $\gamma_p^q$ and $\Gamma_{pq}^{rs}$ are the one-body and two-body reduced density matrices respectively, obtained from CC wave functions. 

Second, it was calculated directly from the Galitskii-Migdal formula \cite{Galitskii1958} using $\Sigma(\omega)$ and $G(\omega)$ obtained from various GFCC variants considered here. In this formalism the 1- and 2-body contributions to the energy are defined as
\begin{equation}
    E_{1b}^{GM} = E_{1b}^{CC} (\gamma(G)) ; \quad E_{2b}^{GM} = \frac{1}{\beta} \sum_{pq} \sum_\omega G_{pq} (\omega) \Sigma_{pq} (\omega), \label{eq:GM-energy}
\end{equation}
where $\gamma(G)=-G(\tau=\beta^{-})$ is obtained from the correlated Green's function. For the orbital-trace we use atomic orbitals (AO) here. Since we are comparing the traced quantities, the difference in the choice of orbital basis between different approaches does not play any role.

We have summarized our results in Table \ref{tab:energycompare}. We expect that GFCCSD and GFCCSDT(2,3) will reproduce E$_{1b}$ and E$_{2b}$, individually from the CCSD wave function calculation, and that GFCCSDT will be comparable to CCSDT wave function results. For O:$p_\sigma$ impurity, both E$_{1b}$ and E$_{2b}$ were recovered with very good accuracy by all the methods considered. This finding can be rationalized from the fact that all these methods were very accurate for evaluating the self-energy of this impurity, as discussed in the previous section. For the O:$p_\pi$ impurity, we observe substantial differences between CCSD(wf) and GFCCSD; both for E$_{1b}$ and E$_{2b}$. When GFCCSDT(2,3) is considered for the O:$p_\pi$ impurity, where 3h-2p/3p-2h projections are used for the the N-1/N+1 particle states, we see a good agreement with CCSD(wf) quantities. We believe this asserts the improved behaviour of GFCCSDT(2,3) over GFCCSD. For GFCCSDT, we hoped that the contribution of the missing projections, that is, of 4h-3p/4p-3h states will be insignificant because they typically appear as a much higher-order perturbative contribution. However, our comparison does not comply with that expectation for O:p$_\pi$ impurity. For Mn:$e_g$ impurity, our observations are very much the same, albeit the numerical values of differences are much smaller.

\begin{table}

\begin{tabular}{|c|c|c|c|}
\hline 
imp & method & E$_{1b}$ & E$_{2b}$\\
\hline 
\hline 
O: p$_\sigma$ & CCSD (wf) & -9.610876 & -0.067456 \\
\cline{2-4} \cline{3-4} \cline{4-4} 
 & CCSDT (wf) & -9.610872 &	-0.067462 \\
\cline{2-4} \cline{3-4} \cline{4-4} 
 & GFCCSD & -9.610168 &	-0.068162 \\
\cline{2-4} \cline{3-4} \cline{4-4} 
 & GFCCSDT(2,3) &  -9.610188 & -0.068143 \\
\cline{2-4} \cline{3-4} \cline{4-4} 
 & GFCCSDT &  -9.610183	& -0.068149 \\
\hline 
O: p$_\pi$ & CCSD (wf) & -5.692181  & -0.162969 \\
\cline{2-4} \cline{3-4} \cline{4-4} 
 & CCSDT (wf) & -5.670236 &	-0.196080 \\
\cline{2-4} \cline{3-4} \cline{4-4} 
 & GFCCSD &  -5.678532 & -0.174392 \\
\cline{2-4} \cline{3-4} \cline{4-4} 
 & GFCCSDT(2,3) &  -5.690790 & -0.161024 \\
\cline{2-4} \cline{3-4} \cline{4-4} 
 & GFCCSDT &  -5.653915	& -0.207156 \\
\hline 
Mn: e$_g$ & CCSD (wf) & -26.130999 & -0.087155 \\
\cline{2-4} \cline{3-4} \cline{4-4} 
 & CCSDT (wf) &  -26.12229 & -0.10047 \\
\cline{2-4} \cline{3-4} \cline{4-4} 
 & GFCCSD & -26.128809 & -0.089666 \\
\cline{2-4} \cline{3-4} \cline{4-4} 
 & GFCCSDT(2,3) & -26.131343 & -0.087726  \\
\cline{2-4} \cline{3-4} \cline{4-4} 
 & GFCCSDT & -26.12050 & -0.10255 \\
\hline 
\end{tabular}
\caption{One-body (E$_{1b}$) and two-body (E$_{2b}$) energies in a.u. evaluated with 1-RDM and 2-RDM from different CC variants and from Galitskii-Migdal formula using $G(\omega)$ and $\Sigma(\omega)$ obtained from different GFCC variants.}
\label{tab:energycompare}
\end{table}

For the self-consistent SEET study with GFCC solver, we have considered MnO solid, which is a prototypical strongly correlated system with AFM ordering. We have studied this system in our previous work \cite{Yeh_SheePRB21} with both the ED and GFCCSD solvers. The computational setup remained the same as in Ref.~\cite{Yeh_SheePRB21}. For MnO, we have observed that updating the GW self-energy via the outer loop self-consistency has very little effect unlike in the case of SrMnO$_3$ solid. Moreover, some of the impurities did not yield correct spectral function with GFCCSD solver in comparison to the ED solver, for example, when O:2p orbitals were considered in the impurity construction. For these two reasons, MnO will be considered as a very good test of GFCC solver performance in the rest of the discussion. 

In our study of MnO with the GFCCSD solver for a case that included O:2p impurity problem, we have observed a spurious peak near the Fermi region of the spectral function (see Ref.~\cite{Yeh_SheePRB21}). The origin of this peak can be traced to the significant discrepancy observed in the GFCCSD self-energy when compared with the ED self-energy. This difference is prominent for the imaginary part of the self-energy.

\begin{figure*}
    \centering
    \begin{tabular}{cc}
    \includegraphics[width=0.45\textwidth]{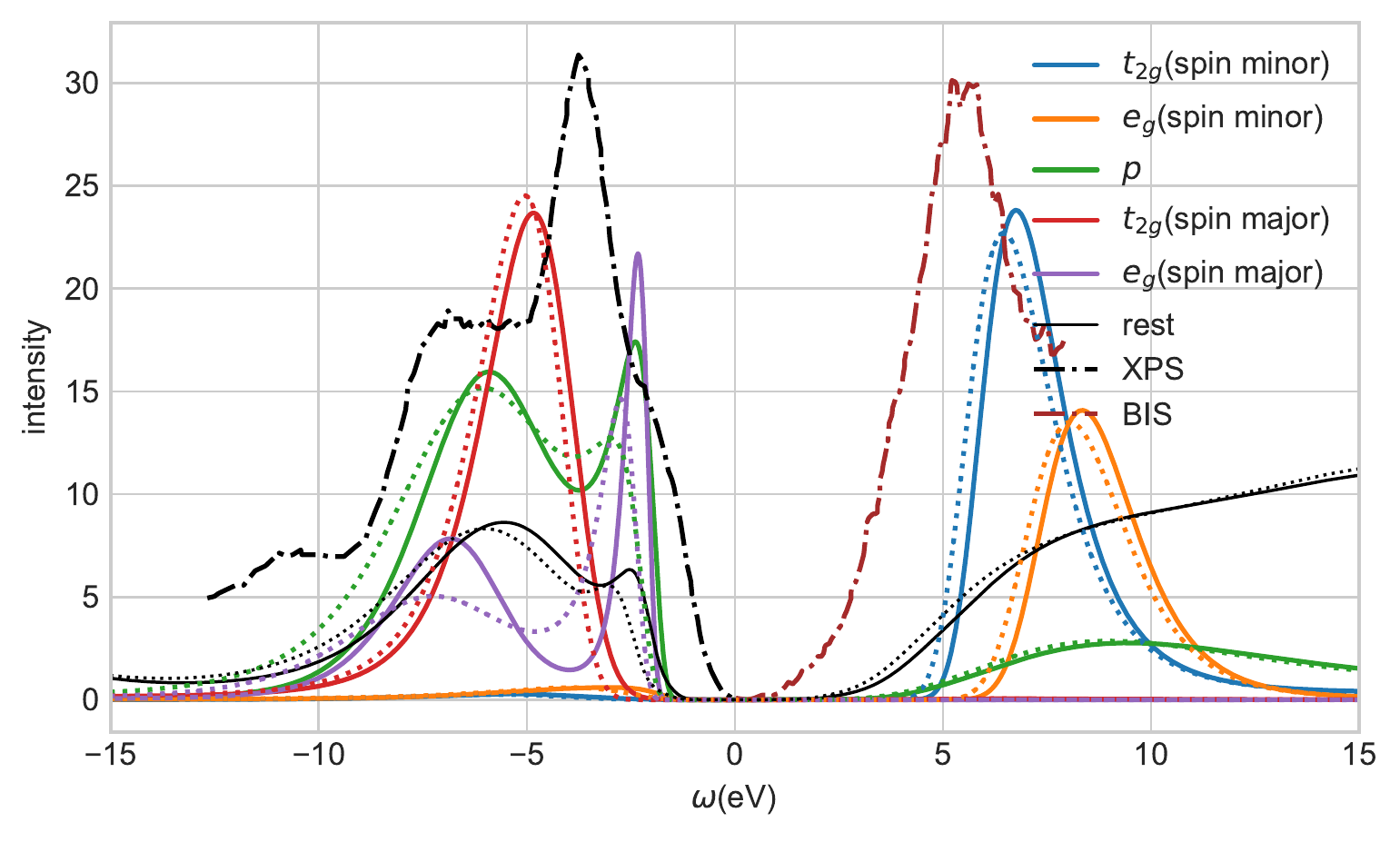} & 
         \includegraphics[width=0.45\textwidth]{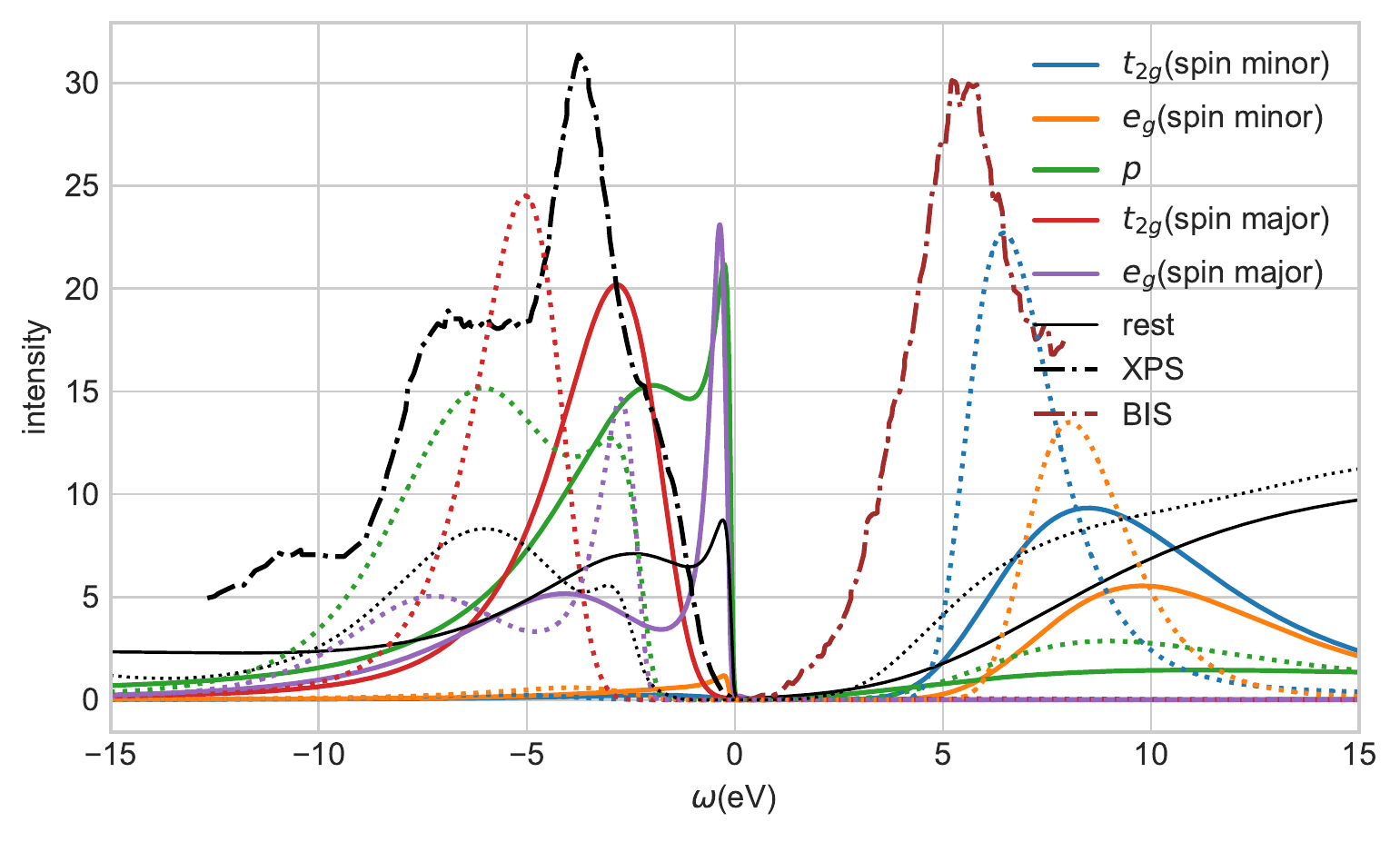} 
    \end{tabular}
    \caption{Orbital resolved spectral function for MnO with GFCCSDT(2,3) solver (left panel) and GFCCSDT solver (right panel). XPS and BIS correspond  to the experimental spectra.}
    \label{fig:scSEET_t_23}
\end{figure*}

When we carried out the full self-consistency calculation with the GFCCSDT(2,3) solver the spurious peak disappeared and the overall band gap in SEET(GW/GFCCSDT(2,3)) spectrum is very close to the spectra for SEET(GW/ED) case. However, the same self-consistent calculation SEET(GW/GFCCSDT) with the GFCCSDT solver failed to remove the spurious peak. We have shown both spectra from SEET(GW/GFCCSDT(2,3)) and SEET(GW/GFCCSDT)  in Fig. \ref{fig:scSEET_t_23}. 

This observation is certainly counterintuitive as GFCCSDT(2,3) is only an approximation to the GFCCSDT method. In the following we will try to provide a plausible explanation of the above outcome.

\begin{figure}
    \centering
    \includegraphics[width=0.47\textwidth]{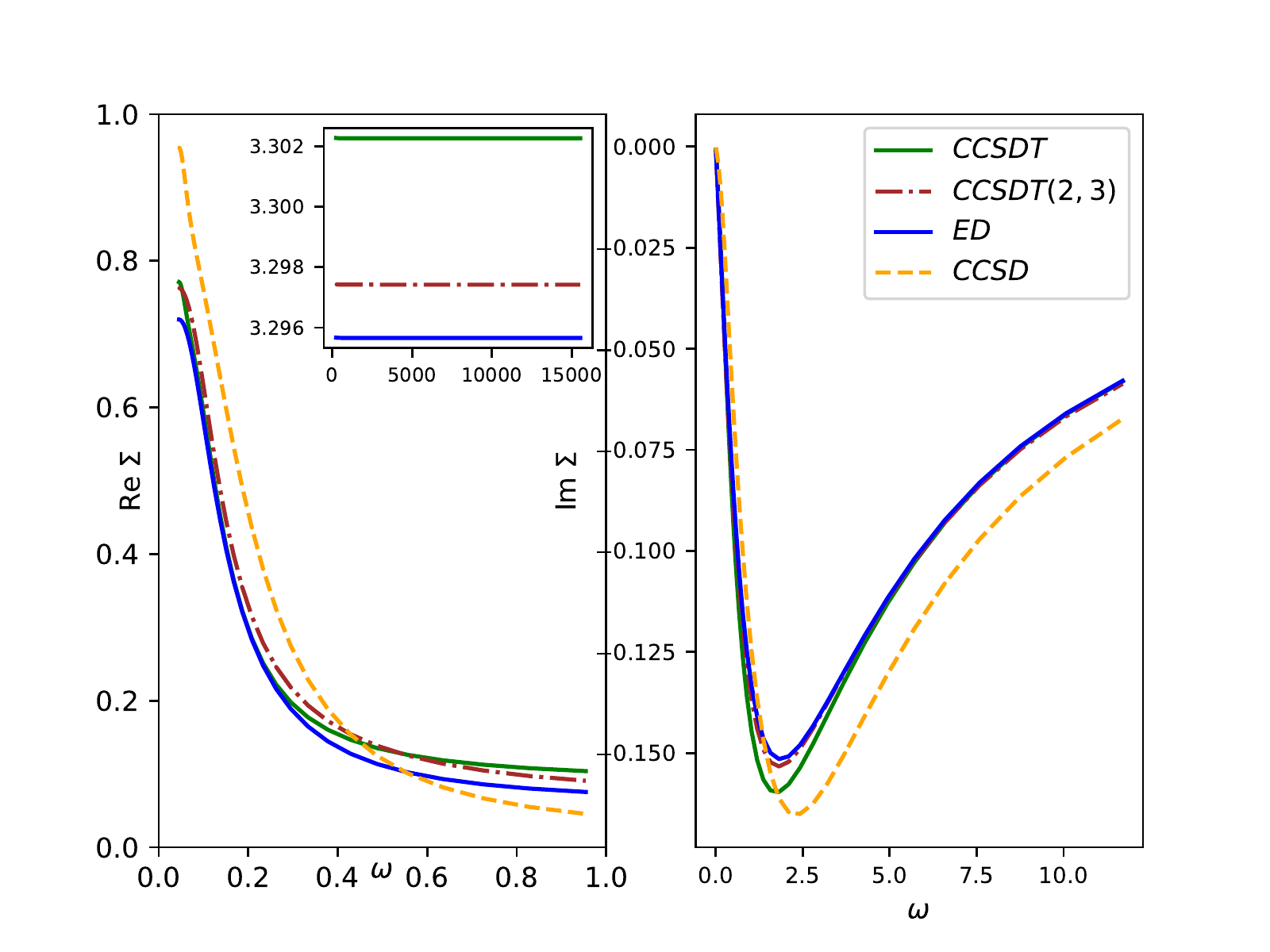}
    \caption{Self-energies obtained from various GFCC impurity solvers for O:2p impurity problem of MnO solid. The inset shows the comparison of the static self-energy from various GFCC variants.}
    \label{fig:mno_p_sigma}
\end{figure}

We plotted the self-energy obtained from ED, GFCCSDT, GFCCSDT(2,3) and GFCCSD solvers for the O:2p impurity problem in Fig.~\ref{fig:mno_p_sigma}. At the low-frequency regime, both GFCCSDT and GFCCSDT(2,3) show much better agreement with the ED self-energy in comparison to the GFCCSD solver. However, if we analyze the Re $\Sigma$ from all the solvers we observe that at high-frequency regime there is significant difference among the solvers - GFCCSDT(2,3) is numerically much closer to what ED predicts, and GFCCSDT is significantly different. We have plotted those quantities in the inset of Fig. \ref{fig:mno_p_sigma}. This large difference in static self-energy causes the chemical potential value in the first iteration of SEET with the GFCCSDT solver to be $\sim$79 mH lower than ED. For GFCCSDT(2,3) this difference is only $\sim$13 mH. We attribute this as the primary reason for the appearance of the spurious peak at the Fermi region.

We furthermore attributed the reason behind this discrepancy to the non self-consistent nature of the GFCC self-energies. To further illustrate this behavior we have evaluated a static contribution $\Sigma_\infty$  to the self-energy in two different ways:
\begin{enumerate}
    \item $\Sigma_\infty = \lim_{\omega \rightarrow \infty} \Sigma(\omega)$
    \item using the first-order self-energy diagrams.
\end{enumerate}

In a self-consistent approach, these two ways of evaluating static contribution to the self-energy will produce the same result (up to the numerical convergence). 
For example, with an ED solver, these two approaches produce the same self-energy. However, both for GFCCSDT and GFCCSDT(2,3), $\Sigma_\infty$ differ quantitatively between these two ways of evaluating it. In our work so far we have taken the first approach mentioned above, where we observe a deviation from ED of $\sim$ 7 mH for GFCCSDT and $\sim$ 2 mH for GFCCSDT(2,3). When this quantity was evaluated with the second approach, surprisingly the deviation for GFCCSDT reduced to 2 mH, but for GFCCSDT(2,3) it increased to 21 mH.   


We have presented a validation of the newly developed variants of the GFCC solver with triple excitations, namely GFCCSDT and GFCCSDT(2,3), on selected impurity problems created for SrMnO$_3$ solid. These benchmark problems demonstrated that the solver involving triples leads to both qualitative and quantitative  improvement of self-energies when compared to the GFCCSD solver. 

Similarly, we have analyzed the 1- and 2-body energy obtained from both CC and GFCC approaches. While for non-self-consistent approaches, such as GFCC, the energies obtained from CC alone and GFCC do not need to agree, we would expect them to be close when sufficient excitations are present in the CC wave function ansatz. Our results show that the addition of triple excitations generally improved this agreement in a substantial manner.

Finally, for MnO solid we have used the GFCCST and GFCCST(2,3) solver in combination with SEET embeding procedure on top of GW calculations. We have observed that SEET(GW/GFCCST(2,3)) gave excellent results reproducing the experimental XPS and BIS spectra. We did not achieve a similar success while employing the GFCCST solver. This unexpected result was explained by analyzing the static part of the GFCCSDT self-energy.

The inclusion of triples correction certainly improves the quality of various dynamical quantities, for example, self-energy, Green’s function etc. obtained. However, the non-self-consistent nature of GFCC precludes the expected systematic improvement. Perhaps a different construction of static self-energy, as alluded in the results section, will improve the final quantities obtained from a self-consistent calculation. In future, we would like to resolve this particular aspect of the theory. Moreover, we can reduce the computational complexity of the triples solver further by employing a so-called AO-driven implementation of the most expensive terms.

A.S., C.-N. Y., B.P., K.K. and D. Z. acknowledge support of the
Center for Scalable, Predictive methods for Excitation
and Correlated phenomena (SPEC), which is funded by
the U.S. Department of Energy (DOE), Office of Science,
Office of Basic Energy Sciences, the Division of Chemical
Sciences, Geosciences, and Biosciences.

\bibliographystyle{apsrev4-1}
\bibliography{cc, group, embed, misc}

\end{document}